\documentclass{article}
\usepackage{amsthm,amsmath,amssymb}
\usepackage{graphics}
\newcommand{\up}[1]{\leavevmode \raise.16ex\hbox{#1}}
\newcommand{\sigln}{{\Sigma\kern-4truept{\up{$\mathbf \mid{ }\,$}}}}
\newcommand{\E}{\mathbb{E}}
\newcommand{\ones}{\mathbf{1}}

\newcommand{\eqind}{\buildrel{\mathbf{D}}\over =}

\begin{document}

\title{The Shrinkage Variance Hotelling $T^2$ Test for Genomic Profiling Studies}

\author{Grant zmirlian}

\newcommand{\Address}{{
  \bigskip
  \footnotesize

  \textsc{Biometry Research Group, Division of Cancer Prevention
          National National Cancer Institute; BG 9609 RM 5E130 MSC 9789
          9609 Medical Center Dr; Bethesda, MD 20892-9789}\par\nopagebreak
  \textit{E-mail address}\texttt{izmirlig@mail.nih.gov}
}}

\maketitle

\begin{abstract}
Designed gene expression micro-array experiments, consisting of several treatment levels
with a number of replicates per level, are analyzed by applying simple tests for group
differences at the per gene level.  The gene level statistics are sorted and a criterion
for selecting important genes which takes into account multiplicity is applied.  A caveat
arises in that true signals (genes truly over or under expressed) are ``competing'' with
fairly large type I error signals.  False positives near the top of a sorted list can
occur when genes having very small fold-change are compensated by small enough variance to
yield a large test statistic.  One of the first attempts around this caveat was the
development of ``significance analysis of micro-arrays (SAM)'', which used a modified
t-type statistic thresholded against its permutation distribution.  The key innovation of
the modified t-statistic was the addition of a constant to the per gene standard errors in
order to stabilize the coefficient of variation of the resulting test statistic.  Since
then, several authors have proposed the use of shrinkage variance estimators in conjunction
with t-type, and more generally, ANOVA type tests at the gene level.  Our new approach
proposes the use of a shrinkage variance Hotelling T-squared statistic in which the per
gene sample covariance matrix is replaced by a shrinkage estimate borrowing strength from
across all genes.  It is demonstrated that the new statistic retains the F-distribution
under the null, with added degrees of freedom in the denominator.  Advantages of this
class of tests are (i) flexibility in that a whole family of hypothesis tests is possible
(ii) the gains of the above-mentioned earlier innovations are enjoyed more fully. This
paper summarizes our results and presents a simulation study benchmarking the new
statistic against another recently proposed statistic.
\end{abstract}

\section{Introduction}
Gene expression microarrays provide a fast and systematic way to identify genes
differentially expressed between two or more experimental groups of samples in a
hypothesis driven study.  These samples and experimental groups could be, for example,
human prostate cancer cell line RNA samples treated with two or three different agents, or
treated with the same agent at differing concentrations. Right now cDNA chips contain on
the order of ten thousand genes, while oligonucleotide arrays contain upwards of twelve
thousand genes.  In the not too distant future entire genome chips will become available.
The consequence is a tremendous savings in time and resources as the per gene expense in
time and resources for the preliminary screening of genes has gone down considerably.
Nonetheless, the considerable cost per array results in experiments that are typically
based upon few replicates.  For example, an experiment consisting of two experimental
conditions might have just three replicates per set of conditions.

While the shift in platfroms from cDNA arrays to oligonucleotide arrays has resulted in
the reduction in various sources of within gene and extra gene variability, the reality is
that there is still a great deal of endemic noise in these sorts of investigations.  Given
the small number of replicates, power is a primary concern.  Albeit, the goal of
statistical analysis in this setting is to arrive at a relatively short list of candidate
genes that warrant further investigation via a more sensitive and specific technique such
as PCR.  The investigator typically has aloted specific resources for the further
investigation of a given number of genes and will request a ``short list'' of the
requesite length.  Therefore, the role of efficiency and power may not be completely
appreciated.  Clearly, however, the goal is to present the best possible list, so that the
role of efficiency and power can now be understood.

A caveat arises in that true signals (genes truly over or under expressed) are
``competing'' with fairly large type I error signals. False positives near the top of a
sorted list can occur when genes having very small fold-change are compensated by small
enough variance to yield a large test statistic.  One of the first attempts around this
caveat was the development of ``significance analysis of micro-arrays'' or (SAM),
\cite{Tusher:2001}, which used a modified t-type statistic thresholded against its
permutation distribution.  The key innovation of the modified t-statistic was the addition
of a constant to the per gene standard errors in order to stabilize the coefficient of
variation of the resulting test statistic.  Since then, (\cite{Wright:2003},
\cite{Lonnstedt:2005}) several authors have proposed the use of shrinkage variance
estimator in conjunction with t-type and more generally, ANOVA type tests at the gene
level.  One advantage of this latter approach is that it doesn't require the computation
of ad-hoc fudge constants.  In the situation under study, e.g. a hypothesis driven
experiment consisting of a small number of experimental groups, a natural model is the per
gene linear model on the appropriate scale, leading to a per gene ANOVA type test of the
null.  Recent work, (\cite{Wright:2003}, \cite{Cui:2005}, \cite{Lonnstedt:2005}),
presented a model in which the per gene residual variance parameters were considered to be
draws from an inverse gamma distribution, resulting in a ``shrinkage variance test'' that
could potentially have gains in efficiency depending on the heterogeneity of the extra
gene variability.  The idea of using a shrinkage estimate of within group variance has
also been persued by others such as \cite{Baldi:2001}, \cite{Lonnstedt:2002}, and
\cite{Menezes:2003}.  One assumption of that model which is often violated in
applications is that the extra gene variability is consistent across experimental
conditions.  In order to circumvent this restrictive assumption, we extend that work to
the multivariate setting arriving now at a whole class of hypothesis tests based upon a
shrinkage variance Hotelling $T^2$.  If there is any appreciable between-group
correlation, this approach constitutes a more efficient use of the scarce data available
per gene data. Furthermore, as we shall point out in this work, the incorporation of a
shrinkage variance/covariance estimator into the usual Hotelling $T^2$ statistic
accomplishes the goals of the earlier innovations to an even greater degree.

\section{Background and Motivation: Designed Gene Expression Micro-Array experiments}
The impetus for this work were two microarray studies with which the authors have been
involved. The first of these was a spotted cDNA array experiment studying the effects of
the isoflavone/phytoestrogen genestein on gene expression in the LnCAP cell line. Several
batches of colonies were treated with either $1\mu$M, $5\mu$M, $25\mu$M, genestein or
control media and allowed to grow for 24 hours.  Messenger RNA (mRNA) isolated from each
of the treated groups was hybridized onto the green channel of a corresponding
micro-array, while mRNA isolated from the control treated colony was hybridized onto the
red channel of each micro-array. This experiment was conducted independently and in
identical fashion on three separate dates. Systematic variability occuring from array to
array and within array were adjusted out in the manner suggested by
\cite{Dudoit:2000}. Within each experimental replicate and for each gene, the log base two
of the ratio of normalized green to red channel expression values were calculated and used
in subsequent analysis. The research questions being investigated were (i) whether there
was differential expression between the green and red channels under treatment with
genestein at any of the three concentrations, and if so (ii) was there a trend in this
effect.

The second study was an oligonucleotide micro-array experiment studying the effects of two
hormones, dehydroepiandrosterone (DHEA) and dihyrotestosterone (DHT), on gene expression
in the LnCAP line.  Again, several batches of colonies were treated with either DHEA, dhT,
or control media and allowed to grow for 24 hours.  mRNA isolated from each of the two
treated colonies as well as from the control treated colony was hybridized onto one of
three corresponding single channel oligonucleotide arrays. The raw image files, in CEL
format, were imported into the R statistical computing platform \cite{Rproj:2004}.  For
each gene, the probe set was summarized into a model based gene expression index
\cite{Li-Wong:2001}, using the Bioconductor suite of add-on libraries for R
\cite{BioC:2004}.  Within each experimental replicate and for each gene, the log base two
of the expression ratios of treatment to control were calculated and used in subsequent
analysis.  The research questions here were (i) whether there was differential expression
between treatment and control under treatment with either hormone, any of the three
conditions, and if so (ii) was there differential expression between the two treatments.

\section{The Shared Hotelling $T^2$ statistic}
As indicated in the introductory remarks above, the new methodologic tool introduced here
is a Hotelling $T^2$ statistic for a variety test of the null which incorporates a
shrinkage estimate of the per gene residual variance.  Suppose that each preprocessed
microarray yields expression levels on each of $G$ genes.  In the type of studies dealt
with here we have a total of $n\times d$ such microarrays arising from $n$ identical
replicates of an experiment having $d$ experimental conditions or ``treatments''.  Here as
is usually the case, the measurements being analyzed will be the log base two of a
treatment to control ratio. For each of the $1\le g\le G$ genes, we consider these
measurements as an i.i.d. sequence of $d$-demensional random variables,
$\{Y_{g,i}\,:\,i=1,2,\ldots,n_g\}$, where we allow the possibility that there may be a
different number of measurements for different genes due to reading errors.  We assume
such missingness is completely at random.  Let $\bar Y_g$ and $S_g$ be the $d$-dimensional
sample mean and unbiased sample covariance matrix corresponding to the sample
$\{Y_{g,i}\,:\,i=1,2,\ldots,n_g\}$.  Denote by $F_{n_1,n_2}$ and $F_{n_1,n_2,\theta}$ the
CDFs corresponding to central and non-central $F-$distributions, respectively, of degrees
$n_1$ and $n_2$, the latter having non-centrality parameter $\theta$. The following
theorem shows that, under an assumed conjugate prior, we can replace the estimated
covariance matrix in the usual Hotelling $T^2$ test with a shrinkage estimate and still
retain the property that the resulting test has an $F$ distribution under the null
hypothesis.

\vfil\eject
\noindent{\bf Theorem 1:}
{\it Suppose that $\min_g n_g>d$ and for a given gene, $g$, that}
\begin{itemize}
\item[1.]{\it conditional upon $\sigln_g$, $\{Y_{g,i}\,:\,i=1,2,\ldots,n_g\}$ is i.i.d. 
  $N_d(\mu, \sigln_g)$},
\item[2.]{\it \{$\sigln_g\,:\,g=1,2,\ldots,G\}$ is i.i.d. ${\rm InvWishart}_d(\nu, \Lambda)$ 
{\it and independant of the above.}}
\item[  ]{{\it Let $T_g^2 = n_g \bar Y_g' \Big(\Lambda + (n_g-1) S_g \Big)^{-1} \bar Y_g$.}}
\end{itemize}
\begin{eqnarray}
&& \kern-7em {\rm\it Then~under~}H0: \mu=0_d,\nonumber\\
&& \kern-7em ShHT_g^2 = \frac{\nu+n_g-2 d-1}{d} T_g^2 {\rm\it ~has~the~} F_{d, \nu+n_g-2 d -1} 
{\rm\it ~distribution.}\label{eqn:ShHT2}
\end{eqnarray}
The model in items 1 and 2 above is called the multivariate normal/inverse Wishart model
in the following.  The above statistic has the potential for fair sized gains in
efficiency.  The most ideal situation occurs when the average (over genes) of the within
gene variability is reasonably small but there is reasonable spread across genes in the
magnitude of this variation.  In such a case, the parameter $\Lambda$ would not add so
much magnitude to the denomenator, while the shape parameter, $\nu$ would gives us extra
degrees of freedom as if we had more replicates per experimental condition.  In reality
there is trade off between these two phenomena, and one checks for gain in efficiency by
comparing with the standard Hotelling $T^2$.

\vfill\eject
Next, we note that, as is the case in the usual Hotelling $T^2$ statistic, a whole family
of statistics arises by applying a linear transformation.  We state this as a corollary to
the above theorem.

\noindent{\bf Corollary 1:}
{\it Assume conditions (1) and (2) above except without any restriction on $d$ and $n_g$
relative to oneanother.  Consider the matrix $M$, which is chosen to be of dimension $q
\times d$ of rank $r < \min_g n_g$.  Then we can replace $\bar Y_g, S_g, \Lambda$ and $d$
by $M \bar Y_g, M S_g M', M \Lambda M'$ and $r$ in the theorem above and the conclusions
still follow.}

The above theorem and its corollary are used to test a variety of null hypotheses, $H_0: M
\mu = 0$ where $\mu = \E Y_1$.  There are three natural choices for $M$.  Call these the
``zero means" contrast, $M_{\rm \mu0}$, the ``equal means" contrast, $M_{\mu{\rm eq}}$,and
the ``no trend" contrast, $M_{{\rm trend}0}$.  Specifically, these are given by: $M_{\rm
\mu0} = I_d$, which requires that $n>d$, $M_{\mu{\rm eq}} = I_d - \frac{1}{n} \ones_d\,
\ones_d'$, which requires that $n>d-1$, and $M_{{\rm trend}0}=\{(u\,u')^{-1}\,u'\}_2,
u=[\ones_d, [0, 1,\ldots,d-1]']$, which requires that $n>1$ and $d>2$.  The application of
these results to testing hypotheses in the analysis of both cDNA and oligonucleotide
arrays will be clearly laid out in the section which follows.

Notice in the definition of the statistics ${\rm ShHT}_g^2$ given above in
\ref{eqn:ShHT2}, the parameter matrix, $\Lambda$, and the shape parameter, $\nu$ arising
in the prior distribution of $\sigln_g$ are assumed to be ``known".  The next result is used
to estimate $\Lambda$ and $\nu$ via maximum likelihood using the data the
$S_g,\,g=1,\ldots,G$ which under our model are i.i.d. draws from the density given below
in \ref{eqn:mgnldstbn}.

\vspace{18truept}
\noindent{\bf Theorem 2:}
{\it Under the conditions of theorem 1, $A_g = (n-1)S_g$ has density function equal to} 
\begin{eqnarray}
f(A) &=& 
\frac{\Gamma_d\left(\frac{\nu+n_g-d-2}{2}\right)}{\Gamma_d\left(\frac{n_g-1}{2}\right)
\Gamma_d\left(\frac{\nu_g-d-1}{2}\right)}
\frac{\left|\Lambda\right|^{(\nu-d-1)/2}\left|A\right|^{(n_g-d-2)/2}}
{\left|\Lambda+A\right|^{(\nu+n_g-d-2)/2}}.\label{eqn:mgnldstbn}
\end{eqnarray}

\section{Other statistics under study}
Here we will be using the notation of theorem 1 and its corollary above.  In addition to
the quantities presented there, we write $Y_g$ for the $n_g d$ dimensional column vector
containing the observations $\{Y_{g,i,k}\,:\,i=1,\ldots,n_g, k=1,\ldots,d\}$ stacked by
replicate within component, and $R_g = (n-1) \sum_{k=1}^{d} S_{g, k, k}$ for the total
within group sum of squares. Notice that within a particular set of distributional
assumptions on $Y_g$ and a particular framework for constructing test statistics, a
variety of hypothesis tests are made possible through the application the appropriate
linear transformation to the data.  That being said, we restrict attention in this portion
of exposition to tests of zero group means. The simplist statistic is the standard
F-statistic, which assumes that the sequence of random vectors
$\{Y_{g,i}\,:\,i=1,\ldots,n_g\}$ is stochastically independent with identical distribution
given by independent normals having component means $\mu_{g,1},\ldots,\mu_{g,d}$ and
common variance $\sigma^2_g$. Under the null hypothesis and under these distributional
assumptions,
\begin{eqnarray}
{\rm UT}_g^2 &=& \frac{d(n_g - 1)}{n_g d}\frac{Y_g'Y_g}{R_g} \label{eqn:UT2}
\end{eqnarray}
has an $f$ distribution with $n_g d$ and $d (n_g - 1)$ degrees of freedom. This standard
$f$-statistic was modified using a univariate empirical bayes estimate of the per-gene
common variance in \cite{Wright:2003}. Similar results and extensions were presented in
\cite{Lonnstedt:2002}, \cite{Cui:2005}, and \cite{Lonnstedt:2005}. The
distributional assumptions required by that method are, conditional upon $\sigma^2_g$,
identical to the above. The difference is that $\sigma^2_g$ has an inverse gamma
distribution with shape parameter $2 s$ and rate parameter $2 r$ which results in
exchangeable dependence among replicates $i$ of the experiment.  Under the null hypothesis
and under these assumptions,
\begin{eqnarray}
{\rm ShUT}_g^2 &=& \frac{2 s + d(n_g - 1)}{n_g d}\frac{Y_g'Y_g}{2 r + R_g} \label{eqn:ShUT2}
\end{eqnarray}
has an $f$ distribution with $n_g d$ and  $2 s + d (n_g - 1)$ degrees of freedom. This is
deduced via an argument almost identical to that in \cite{Wright:2003}.  This statistic is
the univariate analogue of the statistic presented here.

If we assume instead, that the sequence of random vectors $\{Y_{g,i}\,:\,i=1,\ldots,n_g\}$
is stochastically independent with identical distribution multivariate normal with mean
vector components $\mu_{g,1},\ldots,\mu_{g,d}$ and variance covariance matrix $\sigln_g$
then under the null hypothesis and under these distributional assumptions,
\begin{eqnarray}
{\rm HT}_g^2 &=& \frac{n_g - d}{d}\frac{n_g}{n_g-1} \bar Y_g' S_g^{-1} \bar Y_g \label{eqn:HT2}
\end{eqnarray}
has an $f$ distribution with $d$ and $n_g - 1$ degrees of freedom (see, for example, 
\cite{Muirhead:1982}).  

\section{Software: R package SharedHT2}
An R package for conducting analyses using the methods of this paper has been created 
and is available for download at the CRAN website.  One of the most desirable facets 
of this package is that it is entirely coded in C with minimal processing done in R.  
The main function ``EB.Anova'' fits the multivariate normal/inverse Wishart 
model to micro-array data and calculates the per gene $ShHT2$ statistics shown in formula
\ref{eqn:ShHT2} in theorem 1 when the argument ``Var.Struct'' is set to ``general''.
In addition, the same function can be used to fit the normal/inverse gamma model of 
\cite{Wright:2003} and calculate the $ShUT2$ statistics shown in formula \ref{eqn:ShUT2}
in the preceding section by setting the argument ``Var.Struct'' to ``simple''. In both 
cases, the models are fit using maximum likelihood estimation. There is flexibility in 
the choice of hypothesis test via setting the argument ``H0'' to one of the following 
choices. Under the ``general'' variance structure option,
\begin{itemize}
\item[(i)]{if $n > d$, the H0=``zero.means'' null may be tested.}
\item[(ii)]{if $n > d-1$, the H0=``equal.means'' null may be tested.}
\item[(iii)]{if $n > 1$, H0=``no.trend'' null may be tested, but of course this only makes 
sense if $d > 2$.}
\item[(iv)]{The user may also set H0 to an custom contrast matrix of dimension $r\times d$
and of rank $r$.}
\end{itemize}
Under the ``simple'' variance structure option, any of the above null hypotheses may be
tested as long as $n>1$.  By default, ``H0'' is set to ``equal.means''.  The package uses
S3 classes and has several other nice features. For example if the data comes from an affy
experiment and the rows are named after the affy gene identifiers, then a genelist sorted
on p-value can be browsed in the html viewer with links to the Weizmann Institute's
``GeneCards'' database.  Additionally, the simulation study presented in the following
section may be repeated using included functions. It is worth mention here that these
simulations were only made possible by migrating the entire procedure including the loop
over simulation replicates, into C. The interested reader is encouraged to browse the
documentation.

\section{Comparison with other approaches--simulation study}
We conducted a simulation study in order to compare the operating characteristics of the
proposed shared variance Hotelling $T^2$ statistic (${\rm ShHT}^2$) in expression
\ref{eqn:ShHT2} with those of the three other statistics, \ref{eqn:UT2}, \ref{eqn:ShUT2},
\ref{eqn:HT2} that were described in a preceding section. In all cases the test was
relative to the null hypothesis of group means identically zero, with two groups.

The first simulation study was conducted by generating data from the multivariate
normal/inverse Wishart model with $d=2$ groups and $n_g = 3$ replicated observations for
each of G=12625 genes, using values for $\Lambda$ and $\nu$ that were obtained in the
analysis of the oligonucleotide array data (see below for further details). One hundred of
the genes were designated as ``true positives'' by giving them non-zero group specific
means that were chosen in the following way. First, a value of $\theta$ was chosen so that
\[
0.90 = F_{6, 4, 3\theta}(F^{-1}_{6,4}(1-0.0026))
\]
i.e., so that the ${\rm UT}^2$ statistic would have power $90\%$ at a type I error of
0.26\% to reject the null hypothesis of zero group means.  This value of $\theta = 7.5$
was then multiplied by the average per group standard deviation calculated under the
multivariate normal/inverse Wishart model, i.e. $\frac{1}{\nu - 2 d - 2} {\rm
diag}\big[\Lambda\big]$ to arrive at the two group specific means applied identically to
each of the ten designated genes.

In order to study the robustness of the test statistic to lack of model assumptions, a
second simulation study was conducted using a Normal-2 component mixed inverse Wishart
distribution.  Specifically, the data are i.i.d. multivariate normal but the prior
distribution on the random variance/covariance matrix is a mixture of two inverse
Wisharts, having shape parameters $\nu_1$ and $\nu_2$ and common rate matrix $\lambda$.
The mixing proportion, $f$ and shape parameters $\nu_1$ and $\nu_2$ were chose so that the
expected value of $S_g$, the per gene empirical covariance matrix, would remain identical
its value under the multivariate normal/inverse Wishart model used previously,
$\frac{\Lambda}{\nu-2d-1}$. The values used were $f=0.2, \nu_1=18.4067$, and
$\nu_2=6.77542$.  Once again, one hundred genes were designated as ``true positives'' by
assigning means as above.

The simulation results were summarized in two ways. The first method, shown in tables
\ref{tbl:SimMod} and \ref{tbl:SimOth}, used the Benjamini-Hochberg FDR stepdown procedure
to set the significance criterion. In each simulation replicate, the four listed
statistics and corresponding p-values were calculated for each of the 12625 genes.  Next,
for each statistic, the list was sorted on corresponding p-value and the row containing
the largest p-value not exceeding $ (rank) FDR/12625 $ and all rwos above it were marked
significant. The true positive rate was derived as the number of genes called significant
as a proportion of those truely differentially expressed, i.e. 100. The false positive
rate was derived as the number of genes called significant not among those 100.  These
were averaged over simulation replicates yielding emprical true positive rates ($eTPR$)
and empirical false positive rate ($eFPR$). In table \ref{tbl:SimMod} is shown results for
the data simulated from the normal/Inverse Wishart model. The leftmost column is the
nominal false discovery rate, $FDR$, used in setting the significance criterion.  The next
eight columns are the empirical true positive and false positive rates for each of the
four benchmarked statistics.

Results corresponding to data simulated from the multivariate normal/inverse Wishart model
are shown in table \ref{tbl:SimMod}. In the case of the proposed statistic, $ShHT2$, the
$eFPR$ coincides within simulation error with the FDR. That is because the p-values are
derived via the F-distribution listed in theorem 1, which assumes the data arise from a
multivariate normal/inverse Wishart distribution.  Notice as well that the $eTPR$ is quite
high in the 90's at the low FDR of 0.05. The other three statistics benchmarked a clearly
inferior. First, $HT2$, thestandard hotelling $T^2$, is nearly uninformative, displaying
an $eTPR$ of 100\% at all values of $FDR$ with correspondingly high $eFPR$ ranging upwards
from 85\%. The shrinkage variance F-statistic, $ShUT2$, is overly concervative, with
$eFPR$ equal to zero within simulation error and $eTPR$ ranging from 20\% to
50\%. Finally, the ordinary F-statistic, $UT2$, is overly conservative at the lower FDR's
of 5\% and 10\%, but then uninformative at the higher FDR's of 15\%, 20\% and 25\%.

The results corresponding to data simulated from the normal/mixed inverse Wishart model
are shown in \ref{tbl:SimOth}.  The only notable difference relateive to remarks made
above is that control over the FDR is now lost, as the $eFPR$ no longer agrees with the
$FDR$.  Still, if the simulation model can be considered an extreme departure from the
model assumptions then use of the FDR=5\% which gives $eFPR=12\%$ and $eTPR=94\%$ should
be acceptible.

On the other hand one may wish to dispesne with any attempts at controling the false
discovery rate at all, and instead, rely on the statistic's ability to provide a more
informative ordering.  In this case, we simply decide how many genes we wish to call
significant and draw the line there.  For the second method of summarizing the simulation
results the $eTPR$ and $eFPR$ were derived this time using, consecutively, each of the
values of the statistic as the significance criterion.  The results for data obeying model
assumptions are shown in figures 1, and for data not obeying model assumptions in figure
2. It is clear that our proposed statistic, $ShHT2$, outperforms the others when the data
obeys the model assumptions presented in theorem 1.  Although this advantage is attenuated
when the data does not obey model assumptions, there is still a modest advantage.  For
this reason we recommend its use over the one dimensional test, $ShUT2$ and the related
SAM of \cite{Tusher:2001}.

\vfil\eject

Table 1
\begin{center}
\begin{tabular}{|l|| c c | c c | c c | c c|}
\hline
      & \multicolumn{2}{|c|}{ShHT2} & \multicolumn{2}{|c|}{HT2} & \multicolumn{2}{|c|}{ShUT2} & 
        \multicolumn{2}{|c|}{UT2} \\
 FDR  &  TPF     & FPF     & TPF     & FPF    & TPF     & FPF      & TPF     & FPF   \\
\hline\hline
0.05 &	0.929 &	0.045 &	1.00 &	0.857 &	0.204 &	0.00 &	0.014 &	0.00  \\
0.10 & 	0.964 &	0.093 &	1.00 &	0.927 &	0.341 &	0.00 &	0.079 &	0.00  \\
0.15 &	0.976 &	0.141 &	1.00 &	0.951 &	0.424 &	0.00 &	1.00  &	0.992 \\
0.20 & 	0.983 &	0.192 &	1.00 &	0.963 &	0.480 &	0.00 &	1.00  &	0.992 \\
0.25 &	0.987 &	0.242 &	1.00 &	0.970 &	0.526 &	0.00 &	1.00  &	0.992 \label{tbl:SimMod}\\
\hline
\end{tabular}
\end{center}

Table 2
\begin{center}
\begin{tabular}{|l|| c c | c c | c c | c c|}
\hline
      & \multicolumn{2}{|c|}{ShHT2} & \multicolumn{2}{|c|}{HT2} & \multicolumn{2}{|c|}{ShUT2} & 
        \multicolumn{2}{|c|}{UT2} \\
 FDR  &  TPF     & FPF     & TPF     & FPF    & TPF     & FPF      & TPF     & FPF   \\
\hline\hline
0.05  &	0.944 &	0.123 &	1.00 &	0.863 &	0.424 &	0.00 &	0.343 &	0.000 \\
0.10  &	0.968 &	0.223 &	1.00 &	0.929 &	0.556 &	0.00 &	0.593 &	0.000 \\
0.15  &	0.976 &	0.307 &	1.00 &	0.951 &	0.626 &	0.00 &	1.000 &	0.992 \\
0.20  &	0.981 &	0.378 &	1.00 &	0.963 &	0.671 &	0.00 &	1.000 &	0.992 \\
0.25  &	0.985 &	0.442 &	1.00 &	0.970 &	0.706 &	0.00 &	1.000 &	0.992 \label{tbl:SimOth}\\
\hline
\end{tabular}
\end{center}

\section{Application: Two Case Studies}
As mentioned in the introductory section, these techniques were used to analyze two
datasets, the first from a spotted cDNA array experiment and the second from an
oligonucleotide array experiment.  In the first experiment, several colonies of LnCAP
cells were allowed to grow for 24 hours in the presence of either control medium or
$1\mu$M, $5\mu$M, or $25\mu$M of genestein. Messenger RNA (mRNA) isolated from each of the
treated groups was hybridized onto the green channel of a corresponding micro-array, while
mRNA isolated from the control treated colony was hybridized onto the red channel of each
micro-array. This experiment was conducted independently and in identical fashion on three
separate dates. Systematic variability occuring from array to array and within array were
adjusted out in the manner suggested by \cite{Dudoit:2000}. Within each experimental
replicate and for each gene, the log base two of the ratio of normalized green to red
channel expression values were calculated and used in subsequent analysis. Since the group
dimension was $d=3$ and the sample size was $n=3$ then a test of the zero means null using
the $ShHT2$ statistic was not possible.  However, we tested the equal means null using
both the $ShUT2$ statistic and the $ShHT2$ statistic.

The second study was an oligonucleotide micro-array experiment studying the effects of two
hormones, dehydroepiandrosterone (DHEA) and dihyrotestosterone (DHT), on gene expression
in the LnCAP line.  Again, several batches of colonies were treated with either DHEA, dhT,
or control media and allowed to grow for 24 hours.  mRNA isolated from each of the two
treated colonies as well as from the control treated colony was hybridized onto one of
three corresponding single channel oligonucleotide arrays. The raw image files, in CEL
format, were imported into the R statistical computing platform \cite{Rproj:2004}.  For
each gene, the probe set was summarized into a model based gene expression index
\cite{Li-Wong:2001}, using the Bioconductor suite of add-on libraries for R
\cite{BioC:2004}.  Within each experimental replicate and for each gene, the log base two
of the expression ratios of treatment to control were calculated and used in subsequent
analysis.  The research questions here were (i) whether there was differential expression
between treatment and control under treatment with either hormone, any of the three
conditions, and if so (ii) was there differential expression between the two treatments.

\hfil\break
Table \ref{tbl:VEStop4}
\begin{center}
\begin{tabular}{|l|r||c c c c c|}
\hline
  &         Gene  &  dhea &    dht & stat &     p-val &  FDR=0.10 \\
\hline\hline                                                      
1 &  34319\_at     & 1.690 & 4.400  & 273.0 & 1.902e-07 & 7.129e-06\\
2 &  36658\_at     & 2.440 & 2.790  &  90.9 & 8.665e-06 & 1.426e-05\\
3 &  33998\_at     & 0.519 & 0.275  &  85.1 & 1.084e-05 & 2.139e-05\\
4 &  38827\_at     & 1.310 & 1.840  &  64.1 & 2.836e-05 & 2.851e-05\label{tbl:VEStop4}\\
\hline
\end{tabular}
\end{center}

\bibliographystyle{amsplain}
\bibliography{sharedHT2-arXiv}

\section{Appendix Proofs of theorems}
\noindent{\bf Proof of theorem 1:}
In the following, for any symmetric matrix with spectral decomposition $A = Q D Q'$, let
$A^{\frac{1}{2}}_{\rm s} = Q D^{\frac{1}{2}} Q'$ be the symmetric square root of $A$.
$A^{-\frac{1}{2}}_{\rm s}$ is the symmetric square root of $A^{-1}$.  Square root matrices
without the subscript ${\rm s}$ are considered Cholesky square roots, but will not appear
in this manuscript. First, rewrite $T^2$ as follows:
\begin{eqnarray*}
T^2 &=& n \left(\sigln^{-\frac{1}{2}}_{\rm s} \bar Y\right)
\Big(\sigln^{-\frac{1}{2}}_{\rm s} \Lambda \sigln^{-\frac{1}{2}}_{\rm s} +
(n-1) \sigln^{-\frac{1}{2}}_{\rm s} S \sigln^{-\frac{1}{2}}_{\rm s}\Big)^{-1} 
      \left(\sigln^{-\frac{1}{2}}_{\rm s} \bar Y\right)\\
&\eqind& n \left(\sigln^{-\frac{1}{2}}_{\rm s} \bar Y\right)
\Big(\Lambda^{\frac{1}{2}}_{\rm s} \sigln^{-1} \Lambda^{\frac{1}{2}}_{\rm s} +
(n-1) \sigln^{-\frac{1}{2}}_{\rm s} S \sigln^{-\frac{1}{2}}_{\rm s}\Big)^{-1} 
      \left(\sigln^{-\frac{1}{2}}_{\rm s} \bar Y\right),
\end{eqnarray*}
where equality in distribution follows from the fact that because 
$\sigln^{-\frac{1}{2}}_{\rm s} \Lambda \sigln^{-\frac{1}{2}}_{\rm s}$ and
$\Lambda^{\frac{1}{2}}_{\rm s} \sigln^{-1} \Lambda^{\frac{1}{2}}_{\rm s}$ are both
positive definite and symetric, it follows from theorem A9.9 of \cite{Muirhead:1982} that
they are an orthogonal similarity transformation of eachother and since the latter has a
Wishart distribution (see below), equality in distributions follows from the invariance of
the Wishart distribution to orthogonal similarity transformations.

\noindent Next we make the following observations:
\begin{itemize}
\item[1.] {$(n-1) \sigln^{-\frac{1}{2}}_{\rm s} S \sigln^{-\frac{1}{2}}_{\rm s}$ has the 
${\rm Wishart}_d(n-1, I_d)$ distribution and is therefore, independent of $\sigln$.  This is because the 
conditional distribution of $(n-1)S$ given $\sigln$ is ${\rm Wishart}_d(n-1, \sigln)$.}
\item[2.] {$\Lambda^{\frac{1}{2}}_{\rm s} \sigln^{-1} \Lambda^{\frac{1}{2}}_{\rm s}$ has the 
${\rm Wishart}_d(\nu-d-1, I_d)$ distribution, because $\sigln^{-1}$ has the 
${\rm Wishart}_d(\nu-d-1, \Lambda^{-1})$.} 
\end{itemize}
Therefore, the sum, 
\[
V = \Lambda^{\frac{1}{2}}_{\rm s} \sigln^{-1} \Lambda^{\frac{1}{2}}_{\rm s} +
(n-1)\sigln^{-\frac{1}{2}}_{\rm s} S \sigln^{-\frac{1}{2}}_{\rm s}
\]
has the ${\rm Wishart}_d(\nu+n-d-2, I_d)$ distribution.  
Next, put $Z = \sigln^{-\frac{1}{2}}_{\rm s} \bar Y$ and rewrite $T^2$ as
\[
T^2 = n Z'V^{-1}Z = \frac{nZ'Z}{\frac{Z'Z}{Z'V^{-1}Z}}.
\]
Notice that since $Z$ has been rescaled, it is independant of $\sigln$.  Next, because $Z$
is the sample mean, it is independant of the sample covariance matrix, $S$.  Thus $Z$ and
$V$ are independant.  Next, it follows from theorem 3.2.12 of \cite{Muirhead:1982}, the
denominator is distributed $\chi^2_{\nu+n-2 d - 1}$ and independent of the $Z$.  Because
the numerator is $\chi^2_d$, it follows that $T^2 \;\frac{\nu+n-2 d-1}{d}$ has the $F_{d,
\nu+n-2 d-1}$ distribution.

\vspace{18truept}
\noindent{\bf Proof of theorem 2:} 
As stated above, the conditional distribution of
$A=(n-1)S$ given $\sigln$ is ${\rm Wishart}_d(n-1, \sigln)$ which has density:
\begin{eqnarray*}
f^W_{d,n-1,\sigln}(A) = \Gamma_d\left(\frac{n-1}{2}\right)^{-1}
\frac{\left|A\right|^{(n-d-2)/2}}{2^{d(n-1)/2}\left|\sigln\right|^{(n-1)/2}} {\rm
etr}\left(-\frac{1}{2}\sigln^{-1} A \right)
\end{eqnarray*}
while $\sigln$ has the ${\rm InvWishart}_d(\nu, \Lambda)$ distribution, which has density:
\begin{eqnarray*}
f^{W^{-1}}_{d,\nu,\Lambda}(\sigln)=\Gamma_d\left(\frac{\nu-d-1}{2}\right)^{-1}
\frac{\left|\Lambda\right|^{(\nu-d-1)/2}}{2^{d(\nu-d-1)/2}\left|\sigln\right|^{\nu/2}}{\rm
etr}\left(-\frac{1}{2}\sigln^{-1} \Lambda \right)
\end{eqnarray*}
Taking the product of the two above densities and reorganizing factors yeilds:
\begin{eqnarray*}
f^W_{d,n-1,\sigln}(A) f^{W^{-1}}_{d,\nu,\Lambda}(\sigln) \kern-0.75em&=&\kern-0.75em
\Gamma_d\left(\frac{\nu+n-d-2}{2}\right)^{-1}
\frac{\left|\Lambda+A\right|^{ (\nu+n-d-2)/2}}{2^{d(\nu+n-d-2)/2}\left|\sigln\right|^{(\nu+n-1)/2}}{\rm
etr}\left(-\frac{1}{2}\sigln^{-1} (\Lambda + A)\right)\\
&&\frac{\Gamma_d\left(\frac{\nu+n-d-2}{2}\right)}{\Gamma_d\left(\frac{n-1}{2}\right)
\Gamma_d\left(\frac{\nu-d-1}{2}\right)}
\frac{\left|\Lambda\right|^{(\nu-d-1)/2}\left|A\right|^{(n-d-2)/2}}
{\left|\Lambda+A\right|^{(\nu+n-d-2)/2}}.
\end{eqnarray*}
Thus, the posterior distribution of $\sigln$ given $A = (n-1)S$ is \hfil\break
\noindent ${\rm InvWishart}_d(\nu+n -1, \Lambda+(n-1)S)$, and so the distribution of 
$A = (n-1)S$ is the one given in expression \ref{eqn:mgnldstbn}.
\providecommand{\bysame}{\leavevmode\hbox to3em{\hrulefill}\thinspace}
\providecommand{\MR}{\relax\ifhmode\unskip\space\fi MR }
\providecommand{\MRhref}[2]{%
  \href{http://www.ams.org/mathscinet-getitem?mr=#1}{#2}
}
\providecommand{\href}[2]{#2}

\Address

\end{document}